\documentclass[12pt,double-column]{iopart}

\usepackage{graphicx}
\usepackage{color}
\usepackage{float}
\def\bldr{{\bf r}}
\def\bg{{\bf G}}
\def\bp{{\bf r^\prime}}
\def\bpp{{\bf r^{\prime\prime}}}
\def\bppp{{\bf r^{\prime\prime\prime}}}

\def\dd{\rm d}

\begin{document}

\title[Stochastic methods \& Moir\'e impurity states in phosphorene]{Stochastic many-body perturbation theory for Moir\'e states in twisted bilayer phosphorene}

\author{Jacob Brooks$^1$, Guorong Weng $^2$, Stephanie Taylor$^2$, and Vojtech Vlcek$^2$}
\address{$^1$ Department of Chemical Engineering, University of California, Santa Barbara, CA 93106-9510, U.S.A.}
\address{$^2$ Department of Chemistry and Biochemistry, University of California, Santa Barbara, CA 93106-9510, U.S.A.}
\ead{vlcek@ucsb.edu}

\vspace{10pt}
\begin{indented}
\item[]September 2019
\end{indented}
\begin{abstract}
A new implementation of stochastic many-body perturbation theory for periodic 2D systems is presented. The method is used to compute quasiparticle excitations in twisted bilayer phosphorene. Excitation energies are studied using stochastic $G_0W_0$ and partially self-consistent $\bar \Delta GW_0$ approaches. The approach is inexpensive; it is used to study twisted systems with unit cells containing $>2,700$ atoms ($>13,500$ valence electrons), which corresponds to a minimum twisting angle of $\approx 3.1^\circ$. Twisted bilayers exhibit band splitting, increased localization and formation of localized Moir\'e impurity states, as documented by band-structure unfolding. Structural changes in twisted structures lift band degeneracies. Energies of the impurity states vary with the twisting angle due to an interplay between non-local exchange and polarization effects. The mechanisms of quasiparticle energy (de)stabilization due to twisting are likely applicable to a wide range of low-dimensional Moir\'{e} superstructures.
\end{abstract}

\vspace{2pc}

\section{Introduction}

Recently, low-dimensional systems have become a focal point of interest of the physics, chemistry, and materials science communities for their unique (opto)electronic properties \cite{geim2013van,schaibley2016valleytronics,castellanos2016all,mak2016photonics,tan2017recent,manzeli20172d}. The quasiparticle (QP) excitations in these compounds are highly tunable by varying the number of monolayers \cite{mak2011seeing,yan2012tunable,tran2014layer,aziza2017tunable,li2017direct,qiu2017environmental}, their composition \cite{wang2015physical,lin20162d,huser2013quasiparticle,thygesen2017calculating}, and the application of external stimuli \cite{goncalves2009surface,calizo2007effect,peng2014strain,han2014strain}. Regular stacking of monolayers with a finite twist angle results in the formation of Moir\'{e} superstructures which are characterized by a periodic variation of the geometry with large correlation lengths \cite{zhang2017interlayer,yao2018quasicrystalline}. In practice, such structures are associated with the formation of localized impurity states whose energy and spatial distribution is determined by the twisting angle ($\theta$) \cite{kang2013electronic,kang2017moire,sboychakov2015electronic}. Hence, twisting provides a powerful tool to control the quantum many-body interactions. 

Quantitative understanding of QPs in twisted bilayers is hindered by the large system sizes that need to be considered. The localization and spatial separation of the Moir\'{e} impurities increases with decreasing $\theta$, which requires unit cells with thousands of atoms. Density functional theory (DFT) \cite{hohenberg1964inhomogeneous,engel2011density} is an affordable first principles approach that can treat such large systems. However, DFT calculations are limited to the inexpensive (semi)local approximation for electronic exchange and correlation (xc) \cite{engel2011density}. Such methodology suffers from large errors, and it cannot (even in principle) predict QP energies and the fundamental band gaps ($E_g$) \cite{martin2016interacting}. Furthermore, the (semi)local approximation to the xc does not properly account for the non-local electron-electron interactions, i.e., the DFT eigenvalues do not incorporate van der Waals effects that are responsible for the bilayer bonding \cite{engel2011density,klimevs2012perspective}. Prediction of QP energies which incorporate weak interactions require many-body perturbation theory. Conventional implementations of such approaches, however, scale too steeply with system size to treat Moir\'{e} superstructures.

Recently, many-body perturbation theory was formulated using a stochastic sampling approach \cite{neuhauser2012expeditious,neuhauser2014breaking,rabani2015time,neuhauser2017stochastic,vlcek2017stochastic,vlcek2018swift}. Because the number of samples decreases for large systems due to self-averaging, the overall cost of a calculations scales linearly with system size \cite{neuhauser2012expeditious,neuhauser2014breaking,vlcek2018swift,Baer2013}. Hence, random sampling methods enable computations for extremely large systems with thousands of electrons without compromising the accuracy of the QP energies  \cite{neuhauser2014breaking,vlcek2018swift,vlcek_PRM}. 

Up to now, the stochastic approach has been limited to finite systems or 3D periodic solids \cite{neuhauser2014breaking,vlcek2017stochastic,vlcek2018swift,vlcek_PRM,vlvcek2019nonmonotonic}. low-dimensional structures, however, require modified boundary conditions. Furthermore, strongly localized states (such as Moir\'{e} impurity states) are expected to worsen the statistical sampling as seen, for example, in calculations involving localized molecular orbitals \cite{vlcek2017stochastic}. Localized states in periodic systems were not studied by stochastic methods up to now.

\begin{figure}
	\centering
	\includegraphics[width=0.5\textwidth]{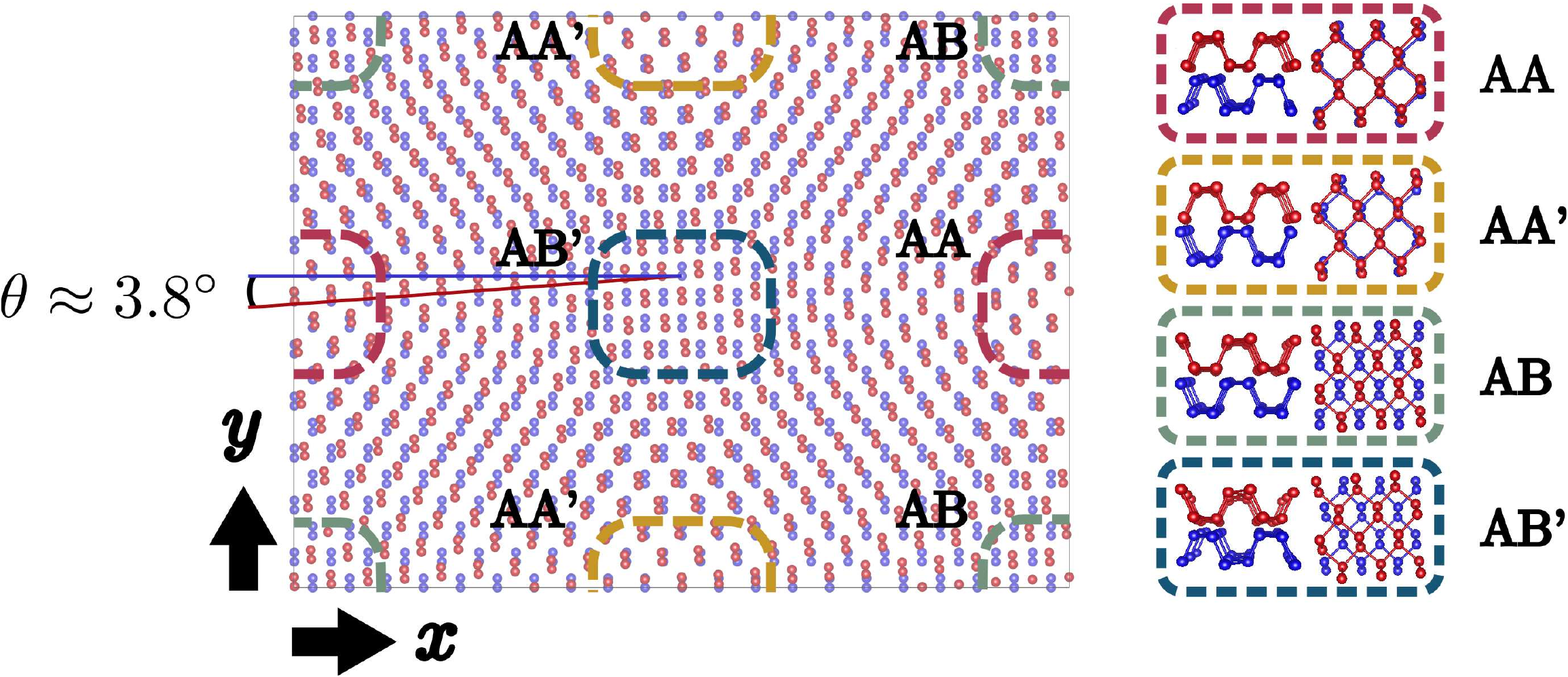}
	\caption{Bilayer phosphorene with twisting angle $\theta\approx3.8^\circ$ is depicted from the $z$-axis perspective (top-down). Black arrows denote the $x$ (zigzag) and $y$ (armchair) directions. The structure is constructed by superimposing two monolayers (atoms are distinguished by red and blue colors) with a rotation along the direction normal to the plane. The mismatch of the two layers causes a finite relative strain. Due to the ridged structure of a monolayer of phosphorene, four characteristic stacking regions (AA, AA', AB, AB') appear. The areas with distinct stacking patterns are distinguished by dashed lines and shown in the insets on the right. The stacking nomenclature is adopted from reference~\cite{kang2017moire}.}
	\label{figure_structure}
\end{figure}

In this paper, we expand the stochastic many-body framework to compute QP energies of twist-induced localized states in low-dimensional semiconductors. We investigate black phosphorene, which shows a large bandgap tunability \cite{qiu2017environmental, tran2014layer}. Twisted bilayers combine a high (opto)electronic anisotropy of individual sheets \cite{liu2014phosphorene} with a range of stable stacking patterns which correspond to distinct polymorphs of multilayer black phosphorene \cite{wu2014tunable}. Hence, the twisted phosphorene structures exhibit a complicated landscape with multiple distinct regions (AA, AA', AB, AB') acting as potential wells for electrons and holes (Figure~\ref{figure_structure}). Recent (semi)local DFT calculations suggested formation of Moir\'{e} impurity states in twisted phosphorene \cite{kang2017moire}, but an investigation into QP states with many-body techniques was elusive up to now.

We compute QP energies in large twisted  phosphorene bilayers with up to $13,000$ valence electrons. The QP energies of the valence and conduction Moir\'{e} states are strongly influenced by many-body effects that are not captured with DFT. We apply a projector-based energy-momentum analysis, which coincides with band structure unfolding for regular supercells \cite{popescu2012extracting,huang2014general}. This analysis is broadly applicable for identification of Moir\'e impurity states. The band structures demonstrate that while twisting only mildly perturbs the low-energy valence states, it causes notable band splitting and QP localization in the near gap region. The Moir\'{e} impurities appear as in-gap states which are well-separated from the rest of the valence bands. This separation decreases for small twisting angles. In contrast, the unoccupied impurities are close to the conduction bands, which are pushed to lower energies with the decreasing $\theta$. The behavior of Moir\'{e} impurities is explained as an interplay between electron localization and non-local electron correlation. The overall conclusions presented here are also applicable to other low-dimensional materials.

The paper is organized as follows: we first review the theoretical approaches and the computational methodology, which is verified at the beginning of the results section. In the following section, we study the effects of structural relaxation and the evolution of QP states with twisting angles. Lastly we present our conclusions.

\section{Theory and Methods}
\subsection{Ground state calculations}\label{sec:ground_state}
We use Kohn-Sham (KS) density functional theory (DFT) \cite{hohenberg1964inhomogeneous,engel2011density,kohn1965self} to obtain the ground state electronic structure, which is the starting point for the many-body calculations (Section~\ref{sec:met_stoch}). The KS Hamiltonian leads to a set of single-particle equations for the eigenstates $\phi$ and corresponding eigenvalues $\varepsilon^{KS}$:
\begin{equation}
\left[ -\frac{1}{2}\nabla^2 + V_{ext}(\bldr) + V_H (\bldr)+ V_{xc}(\bldr) \right] \phi(\bldr) = \varepsilon^{KS} \phi(\bldr),
\end{equation}
The first term in brackets represents the kinetic energy operator, the remaining terms correspond to the external, Hartree and exchange-correlation (xc) potentials. In the absence of external fields, $V_{ext}$ is the potential of nuclei; the Hartree term is the potential due to the total electron density $n(\bldr)$:
\begin{equation}
V_H (\bldr) = \int \nu(\bldr,\bp) n(\bp) \dd \bp, 
\end{equation}
where $\nu(\bldr, \bp)$ is the Coulomb kernel
\begin{equation}\label{eq:Coulomb_kernel}
\nu(\bldr, \bp) = \frac{1}{\left| \bldr - \bp\right|}.
\end{equation}
The xc term is a local mean-field potential, which, in principle, embodies all ground-state electron-electron interactions but is approximated in practice.

For 2D systems, the Coulomb kernel (Eq.~\ref{eq:Coulomb_kernel}) is modified so that the direction perpendicular to the surface of the slab ($z$) is treated aperiodically by truncating $\nu$ in momentum space \cite{spataru2004quasiparticle,rozzi2006exact}. The Coulomb kernel thus depends on the components of the momentum vector perpendicular to and in-plane of the 2D system, denoted $k_{z}$ and $k_{xy}$ \cite{rozzi2006exact}:
\begin{equation}\label{eq:coulomb_cut}
\nu(k_{z}, k_{xy}) =  \left\{  \begin{array}{r@{\quad}cr} 
\frac{4\pi}{k^2} \left[ 1+e^{-k_{xy} R} (\frac{k_{z}}{k_{xy}} \mathrm{sin}(k_{z} R) - \mathrm{cos}(k_{z} R)) \right]& \forall k_{z} \land k_{xy} > 0 \\  
\frac{4\pi}{k^2} \left[ 1-k_{z} R \mathrm{sin}(k_{z} R) - \mathrm{cos}(k_{z} R)  \right]&  k_{z} \neq 0 \land k_{xy} = 0 \\  
-2\pi R^2 &     k_{z} = 0 \land k_{xy} = 0 .
\end{array}\right.
\end{equation}
Here $k^2 = {k_{z}^2 + k_{xy}^2}$ and $R$ is half of the simulation cell dimension in the $z$-direction. The Coulomb kernel cutoff is applied in $V_{ext}$ and $V_H$ terms.

The DFT calculations are performed on a real-space grid with Troullier-Martins pseudopotentials ~\cite{TroullierMartins1991}. The exchange-correlation interaction is described by the PBE functional \cite{perdew1996generalized}. In all calculations, we use a $22\, E_h$ kinetic energy cutoff, which yields DFT eigenvalues converged to $<5$~meV. The grid spacing is $0.50\pm0.04~a_0$; the small variation is due to changes of the unit cell dimensions with relaxation of different structures. A vacuum layer of $30~a_0$ above and below the 2D system along the $z$-direction is sufficient and leads to $<1$~meV errors in $\varepsilon^{KS}$. Overall, the Kohn-Sham eigenvalues are converged to $<10$~meV. 

The real-space implementation was verified against the plane wave Quantum Espresso (QE) code \cite{giannozzi2009quantum}. In the QE calculations, we employed an identical set of norm-conserving pseudopotentials. The Brillouin zone was sampled by a 10$\times$8$\times$1 Monkhorst-Pack grid \cite{monkhorst1976special}. We applied kinetic energy and density cutoffs of 25~$E_h$ and 40~$E_h$. The QE code adopts a different treatment of the 2D periodic boundary conditions \cite{otani2006first,sohier2017density}, yet the agreement with our real-space code is excellent. The difference between our implementation and QE results is $<1$~meV for the band edge states (cf. the discussion in Section~\ref{sec:res_twist} and Appendix B).

The phosphorene monolayer, bilayer and twisted bilayer with $\theta \approx 8.0^\circ$ were optimized in the QE code; van der Waals interactions were treated by the Tkatchenko-Scheffler total energy corrections \cite{tkatchenko2009accurate}. The relaxation of the cell parameters and ionic positions was performed until each component of the residual force vector for each atom was below $5\times10^{-4} E_h/a_0^3$.

\subsection{Quasiparticle energy calculations}\label{sec:met_stoch}
The QP energies are computed via many-body perturbation theory \cite{martin2016interacting}, in which a dynamical and non-local self-energy operator, $\Sigma(\bldr, \bp, t)$, captures the electron-electron interactions. In practical calculations, the self-energy is constructed from a perturbation expansion which has to be approximated; we use the widely successful $GW$ formulation \cite{martin2016interacting,hedin1965new,hybertsen1986electron,aryasetiawan1998gw,hedin1999}:
\begin{equation}\label{eqGW}
\Sigma (\bldr, \bp, t) = G(\bldr, \bp, t) W(\bldr, \bp, t^+),
\end{equation}
where $G$ is the QP Green's function, and $W$ is the screened Coulomb potential \cite{martin2016interacting,aryasetiawan1998gw,hedin1999}. The time argument $t^+$ is infinitesimally after $t$ to guarantee the correct time-ordering.
The $GW$ expression is further approximated by neglecting the self-consistency and the QP energy is obtained by a ``one-shot'' correction, conventionally denoted as $G_0W_0$. Specifically, for the $i^{\rm th}$ KS eigenstate: 
\begin{equation}\label{eqp_eq}
\varepsilon_i^{QP} = \varepsilon_i^{KS} - {V}_{xc, i} + \Sigma_i\left(\varepsilon^{QP}\right),
\end{equation}
where ${V}_{xc, i}$ and $\Sigma_i$ are the expectation values of the xc potential and the self-energy for state $\phi_i$. The self-energy is expressed in the frequency domain, and should be evaluated at the frequency corresponding to the QP energy. 

In practice, we decompose $\Sigma$ to a sum of static and dynamical (frequency-dependent) components, $\Sigma_X$ and $\Sigma_P$, which represent the exchange and polarization self-energies. The expectation values of the exchange term for a KS eigenstate $\phi$ is: 
\begin{equation}\label{eq:X}
\Sigma_X = -\sum_j^{N_{occ}} \int \int \phi^*\left(\bldr\right) \phi_j\left(\bldr\right) \nu(\bldr, \bp) \phi^*_j\left(\bp\right)  \phi\left(\bp\right) \dd \bldr \dd \bp,
\end{equation}
where the sum extends over all $N_{occ}$ occupied states. The polarization self-energy represents a potential due to the induced charge density; its expectation value in the time domain is: \cite{vlcek2017stochastic,vlcek2018swift,vlcek2018simple}
\begin{equation}\label{eq:P}
\Sigma_P\left(t\right) = \int \int \int \int  \phi^*(\bldr) G(\bldr, \bp, t) \nu(\bldr, \bpp) \chi(\bpp, \bppp, t^+) \nu(\bppp, \bp)  \phi(\bp)\, \dd \bldr \dd \bp \dd \bpp \dd \bppp
\end{equation}
where $\chi$ is the \emph{reducible} polarizability \cite{fetter2003quantum}. The self-energy and the polarizability are time-ordered quantities \cite{vlcek2017stochastic,vlcek2018swift,vlcek2018simple}. The time and frequency-dependent representations of $\Sigma_P$ are related by Fourier transformation. 

The $G_0W_0$ method yields QP energies and fundamental band gaps in good agreement with experiments \cite{martin2016interacting,hybertsen1986electron,aryasetiawan1998gw,hedin1999}, though self-consistent treatment is often necessary to achieve this \cite{martin2016interacting,van2006quasiparticle,shishkin2007accurate,caruso2012unified}. Widespread application of $GW$, however, has been hindered by its computational cost which scales as $N^4$ with number of electrons \cite{deslippe2012berkeleygw,nguyen2012improving,pham2013g}. This limitation has been recently overcome by the stochastic $G_0W_0$ formulation \cite{neuhauser2014breaking,vlcek2017stochastic,vlcek2018swift}, a statistical approach in which the expectation values of the self-energy are sampled using random vectors in the Hilbert space. This method leads to substantial computational savings and allows performing many-body calculations in a linear scaling fashion \cite{neuhauser2014breaking,vlcek2018swift}. The time-domain formulation is further combined with partial self-consistency at no additional cost  \cite{vlcek2018simple}. 

In the stochastic formulation, the Green's function is decomposed into a set of random vectors $\zeta$. Two additional sets of stochastic vectors are used to characterize the polarizability and to perform time ordering using the sparse stochastic compression technique \cite{vlcek2018swift}. Each vector represents the entire occupied/unoccupied space, which is conventionally described by the KS orbitals from the ground-state calculation. In practice, the expectation value of the self-energy becomes a statistical estimator. Such reformulation is only exact in the limit of an infinite number of stochastic vectors, therefore a finite number of random states results in a statistical error. However, only a small number of stochastic states is usually required to converge the error below an acceptable threshold. The convergence is discussed in more detail in the following paragraph and in Section~\ref{sec:convergence}.

In this paper, we use a modified version of the StochasticGW code \cite{vlcek2018swift} with Coulomb kernel cutoff for 2D periodic systems (Eq.~\ref{eq:coulomb_cut}). We employ $20,000$ fragmented stochastic bases \cite{vlcek2018swift}. The screened Coulomb potential is sampled by $8$ stochastic orbitals per each stochastic sampling of the Green's function, similarly to a previous study of phosphorene \cite{vlcek_PRM}. The time propagation is performed using the random-phase approximation with a propagation time of $100$ atomic units. The total number of stochastic samples $N_\zeta$ is varied to reach a designated error. The convergence of the QP energies with the number of stochastic vectors $N_\zeta$ is discussed in Section~\ref{sec:convergence}. The QP energies of twisted bilayers in Section~\ref{sec:res_twist} were computed at the $G_0W_0$ level as well as with a simplified self-consistency $\bar{\Delta} GW_0$, in which the Green's function is updated as detailed in \cite{vlcek2018simple}.

\section{Results}

\subsection{Convergence of the quasiparticle gaps}\label{sec:convergence}

In the stochastic formulation, the QP energies are obtained by Eq.~\ref{eqp_eq}, with the expectation values of the self-energy evaluated by statistical sampling. The statistical error in $\Sigma$ is governed by the number of stochastic orbitals $\zeta$ used in the decomposition of the Green's function (see Section~\ref{sec:met_stoch}). In periodic systems, the band gaps tend to converge quickly, with several hundred $\zeta$ vectors usually being sufficient. The fluctuation in the monolayer and bilayer is similar; $N_\zeta\approx 200$ is sufficient for the band gaps of the largest systems studied to a statistical error of $25$~meV.

Next, we turn to convergence with respect to simulation cell sizes. In the ground-state KS DFT calculations, even small supercells (with $\sim 100$ atoms) yield converged total energies and eigenvalues. However, the convergence of the many-body calculations is different. The non-locality of the $GW$ self-energy often requires huge cells to converge the quasiparticle energies.
		
In our implementation, the periodic system is treated by a supercell that has to be larger than the characteristic electron-electron correlation length. This real-space supercell approach is equivalent to the Brillouin-zone sampling using a regular mesh of $k$ points. For systems with strong screening, such as 3D periodic semiconductors, the convergence with the system size is usually rapid \cite{vlcek2018swift,vlcek_PRM}. In free-standing 2D semiconductors, the electron-electron interaction is screened much less (particularly in the direction perpendicular to the surface) \cite{qiu2017environmental,berkelbach2013theory,chernikov2014exciton,qiu2016screening,trolle2017model}. Hence, the characteristic distance of electron-electron interactions (and the simulation cell dimensions) are longer in 2D than in 3D systems.
		
\begin{figure}
	\centering
	\includegraphics[width=0.5\textwidth]{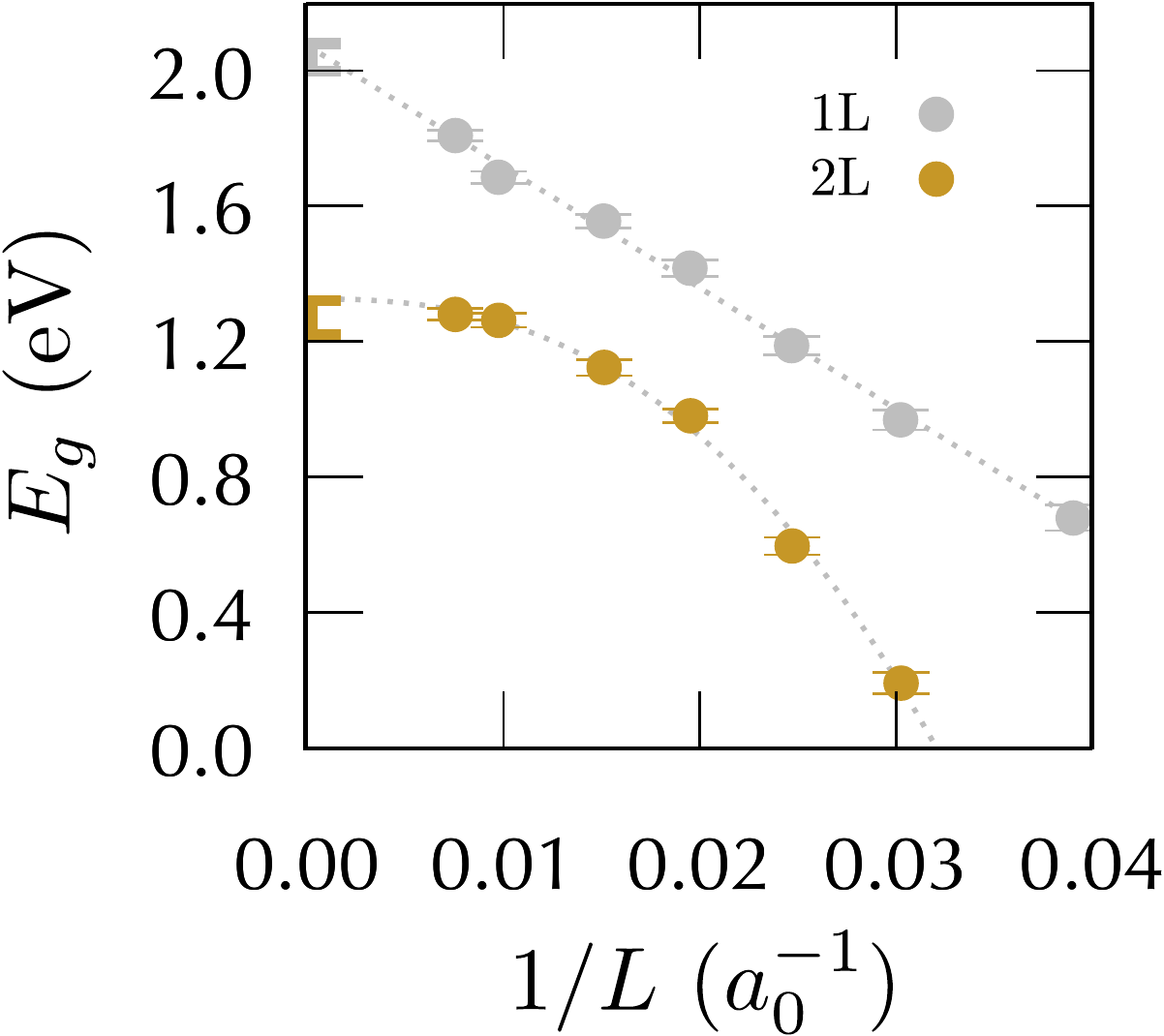}
	\caption{Convergence of the QP band gap with respect to the inverse characteristic length of the monolayer and bilayer supercells. Error bars represent the statistical error due to stochastic implementation. The vertical-axis brackets represent literature values for monolayer and bilayer (see Table~\ref{table_gaps}).}\label{fig:converge}
\end{figure}

For monolayer phosphorene, extremely large supercells are needed. We considered seven systems with sizes up to 1,280 atoms. The values of the QP band gaps (Figure~\ref{fig:converge}) change approximately linearly with the inverse of a characteristic length $L=\sqrt{N_x\cdot a\times N_y \cdot b}$, where $a$ and $b$ are the lattice parameters and $N_x$ and $N_y$ are the number of cells in a supercell along the $x$ and $y$ directions. $N_x$ and $N_y$ were chosen to make the supercells approximately square. However, even the largest monolayer system with $N_x \times N_y = 20\times16$ is not converged, as illustrated in Figure~\ref{fig:converge}. By linear extrapolation, we estimate that $E_g\approx 2.07\pm0.03$~eV for $L\to\infty$. This value is in excellent agreement with the previous $G_0W_0$ estimates, which range between 2.0 and 2.1 eV \cite{tran2014layer,liang2014electronic}.

The convergence for bilayer phosphorene is much faster. We consider six systems with up to 2560 atoms. Their fundamental band gaps (Figure~\ref{fig:converge}) start to converge for supercells larger than $10\times8$ unit cells. For $ N_x \times N_y = 16\times12$ and $20\times16$, the band gaps are almost identical: $1.26\pm 0.05$ and $1.28\pm 0.04$. By fitting a simple power function (shown by a dashed line in Figure~\ref{fig:converge}), we obtain an extrapolated value of $1.31\pm0.03$~eV. These results are in excellent agreement with the previous $G_0W_0$ calculations, which reported values between 1.22 and 1.32 eV \cite{qiu2017environmental,tran2014layer}.
		
Finally, we test the convergence of $E_g$ for a twisted bilayer phosphorene with $\theta\approx 8.0^\circ$, which is commensurate with the $10\times5$ bilayer supercell. Here, the band edge states are localized (discussed below) and the stochastic fluctuations increase by $\sim30\%$. For the 8.0$^\circ$ twisted cell (as for all twisted systems considered in the rest of this paper), $N_\zeta = 400$ is sufficient to converge the gaps to a stochastic uncertainty of $\le 30$~meV. 

The convergence pattern with the cell size is similar as in the bilayer phosphorene. We compared the results for a single $\theta\approx8.0^\circ$ unit cell and a $2\times2$ supercell (i.e., systems with 404 and 1,616 atoms). The QP gap follows the same convergence trend as the bilayer, with a constant difference of $\sim0.36$~eV. Based on extrapolation for the bilayer, the deviation from an $L\to \infty$ limit is $<0.03$~eV for the 1,616 atoms supercell. The size of this error is comparable to the stochastic uncertainty. 
		
To study the effects of structural relaxation (Section~\ref{sec:relax}), we employ a $2\times2$ supercell  of the $\theta\approx8^\circ$ system. To investigate the effect of twist-induced localization (section \ref{sec:relax}), we consider a $2\times2$ supercell of the $8.0^\circ$ twisted bilayer and single unit cells for $\theta \le 5^\circ$. Hence, all the cells considered have $1/L< 0.012~a_0^{-1}$, which is near to the fully converged limit (based on the $E_g$ extrapolation for the bilayer -- see Figure~\ref{fig:converge}). We estimate the total errors to be $<80$~meV.

\subsection{Quasiparticle energies and twist-induced structural changes}\label{sec:relax}
		
A non-zero twisting angle is associated with a Moir\'{e} superstructure characterized by regions with different local stacking orders. Optimization with simple force-fields results in bulging and corrugation of the bilayer related to the local stacking \cite{kang2017moire}, however, our calculation with a state-of-the-art reactive force field \cite{xiao2017development} did not reproduce this result. To avoid potential errors due to a particular force-field parametrization, we use first principles geometry optimization to study the qualitative and quantitative effects of structural relaxation on the band edge states. Here, we investigate the $\theta\approx8.0^\circ$ system, which has a unit cell small enough that the first-principles geometry optimization can be performed (as detailed in Section~\ref{sec:ground_state}).

\begin{figure}
	\centering
	\includegraphics[width=0.5\textwidth]{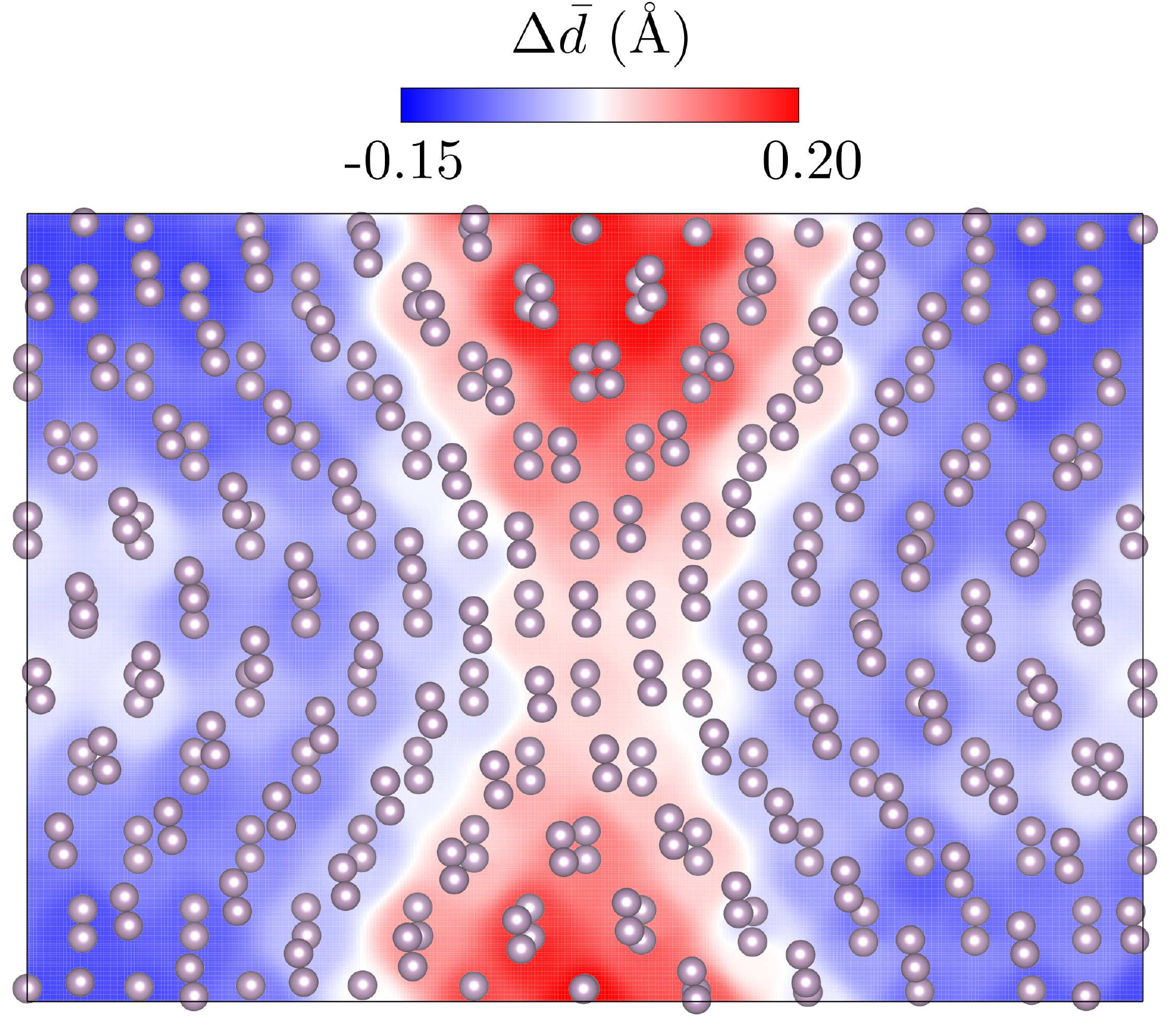}
	\caption{Structure of the relaxed $\theta\approx8^\circ{}$ unit cell is shown along the $z$-direction; the underlying coloring$^\ddagger$\phantom{\footnote{}}depicts the local difference from the average interlayer distance $\bar d = 3.48$~\AA. The interlayer distances within each stacking configuration are provided in Appendix A.}
	\label{fig:T8_relax_heat}
\end{figure}
\footnotetext{The mesh was made with a $1000\times1000$ rectangular grid with the value of each grid space interpolated by an average of the input data weighted by the inverse fourth power of its distance to the grid space.}

\begin{figure}
	\centering
	\includegraphics[width=0.5\textwidth]{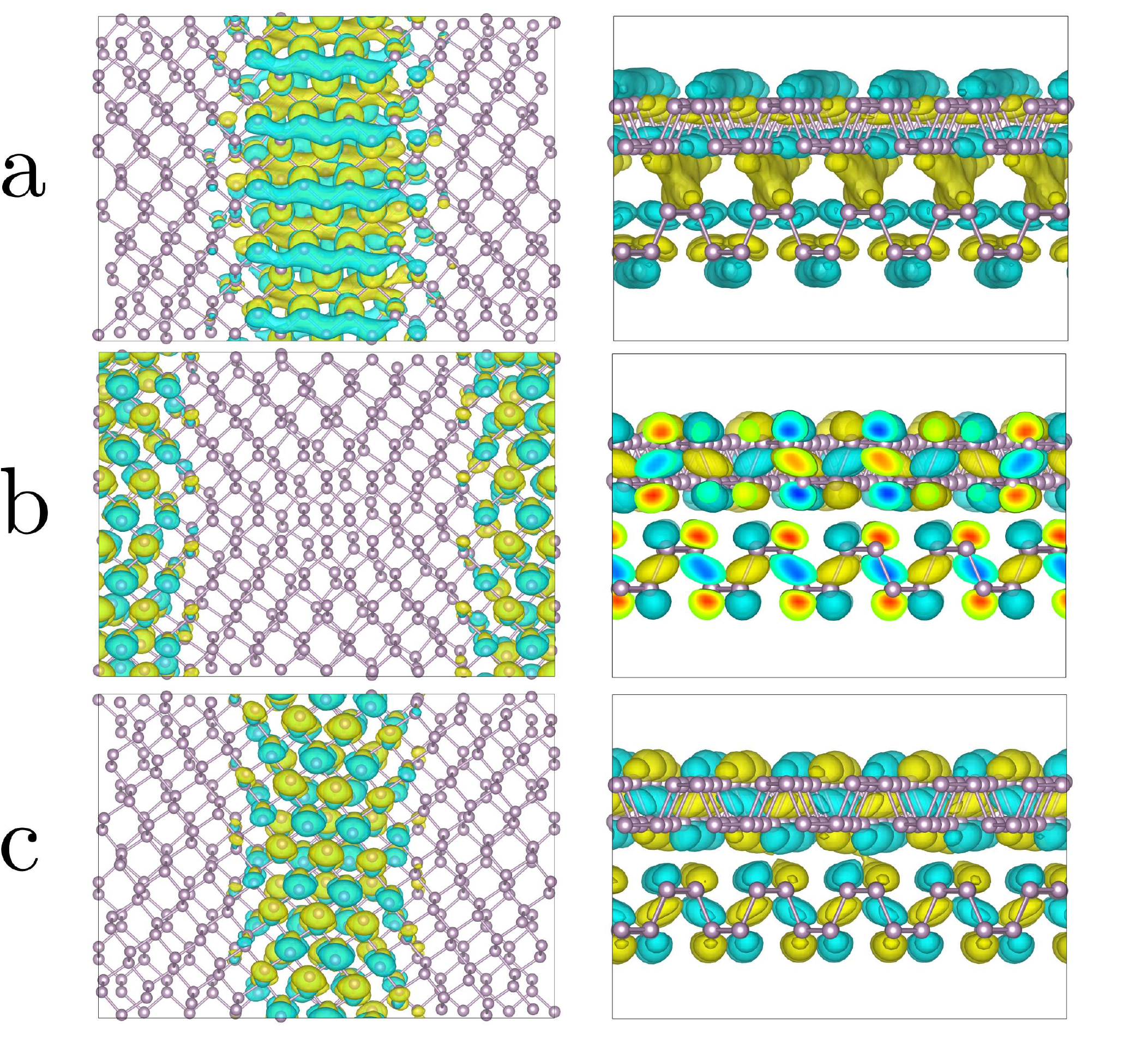}
	\caption{Left and right columns show the isosurface of the band edge states along the $z$ and $x$ directions. The yellow and blue colors denote different wave-function phases. All the states are confined along the $x$-direction. Panel a shows the conduction impurity states bridging the AA'--AB' region; it is localized in the interlayer region of the twisted bilayer. Panels b and c show the isosurfaces of the valence impurities which have a nodal plane in the interlayer region. For planar systems the two valence impurities are degenerate in energy.}
	\label{fig:T8_relax}
\end{figure}
		
We start from a planar structure based on the $0^\circ$-bilayer geometry. After optimization, the average interlayer distance $\bar d$ is increased by $6\%$ to $\bar d = 3.48$~\AA. Large deviations from $\bar d$ are observed in the AA' and AB stacking regions (Figure~\ref{figure_structure}), which correspond to arrangements with the shortest and largest separation of phosphorus atoms in each layer. After the relaxation, the atoms are displaced mostly along the $z$-axis, illustrated in Figure~\ref{fig:T8_relax_heat}.
		
The valence band maximum (VBM) and conduction band minimum (CBM) states are shown in Figure~\ref{fig:T8_relax} for the relaxed structure. The band edge states are strongly confined along the $x$-axis and reflect the topography of the relaxed bilayer surface: the VBM state is localized along the line connecting AA--AB regions, while the CBM state connects the AA'--AB' areas (Figure~\ref{figure_structure}). Besides their spatial distribution, the two states qualitatively differ by the presence/absence of the horizontal nodal plane between the top and bottom phosphorene layer. The VBM and CBM states are distinctly affected by the variation of the average interlayer distance ($\bar d$). The CBM state is preferentially localized in the interlayer area and, hence, it depends on $\bar d$ more as discussed below. 
		
The $G_0W_0$ QP gap of the fully relaxed twisted bilayer is $1.00\pm0.03$ eV, i.e., $27\%$ smaller than in the $0^\circ$-bilayer (cf.~Table~\ref{table_gaps} and Section~\ref{sec:convergence}). Two effects are responsible for the changes in $E_g$: First, the finite $\theta$-angle leads to state localization. The spatial distribution of the VBM and CBM states is affected by the twisting angle, but it is insensitive to the presence of bulging and corrugation. Second, $\bar d$ increases and opens the band gap. We compared two \emph{planar} twisted bilayers with $\bar d = 3.29$~\AA\, and $3.48$~\AA\, and found that their band gaps were $0.78\pm0.03$ and $0.91\pm0.03$~eV. The smaller gap corresponds to the interlayer distance in relaxed $0^\circ$-bilayer. We conclude that the presence of localized Moir\'{e} impurity states together with increased $\bar d$ are the primary causes of band gap changes. 
		
The remaining difference between the gaps of the fully relaxed ($1.00\pm0.03$~eV) and the planar system ($0.91\pm0.03$~eV) is due to bulging and corrugation in the AA' and AB' stacking region (Figure~\ref{figure_structure}) and slight variations in lattice parameters. The largest interlayer distance is $3.72$~\AA, which corresponds to the AA' stacking region (see Figure~\ref{figure_structure}). Since the CBM state is localized between the two monolayers (Figure~\ref{fig:T8_relax}), it is sensitive to the interlayer distance variation. Indeed, if the $\bar d$ of a planar twisted bilayer is increased to 3.72~\AA, the band gap increases to $1.04\pm0.02$~eV. In contrast, the valence impurity energy does not change with $\bar d$ increase. Hence, the result for enlarged $\bar d$ is in excellent agreement with the fully relaxed system despite the absence of bulging and corrugation. The remaining small difference ($\sim 0.04$~eV) is primarily due to the increased lattice parameters of the relaxed twisted bilayer: $a$ changes by $\approx0.3$\% and $b$ by $\approx0.9$\%.
		
Finally, we note that the presence of corrugation lifts degeneracy of occupied states. In planar twisted bilayer, the VBM is doubly degenerate (energy splitting is $<0.02$~eV); the two states connect the AA--AB and AA'--AB' areas (Figure~\ref{fig:T8_relax}). The VBM state degeneracy is not present in the underlying DFT band structure, i.e., it is a result of the many-body treatment.  Hence, corrugation and local distortions affect state ordering, but do not represent significant contributions to $E_g$. 

In the rest of this study, we consider only changes that stem purely from the variation of $\theta$. All the structures are planar and constructed with $\bar d = 3.48$~\AA\, (i.e., the $\bar d$ of the relaxed $8.0^\circ$-degree bilayer). In all structures studied, a small strain ( $<1\%$, see Table.~\ref{table_gaps}) is present due to the mismatch between the upper and lower monolayers.

\subsection{Evolution of Moir\'e impurity states}\label{sec:res_twist}
		
\begin{figure}
	\centering
	\includegraphics[width=0.5\textwidth]{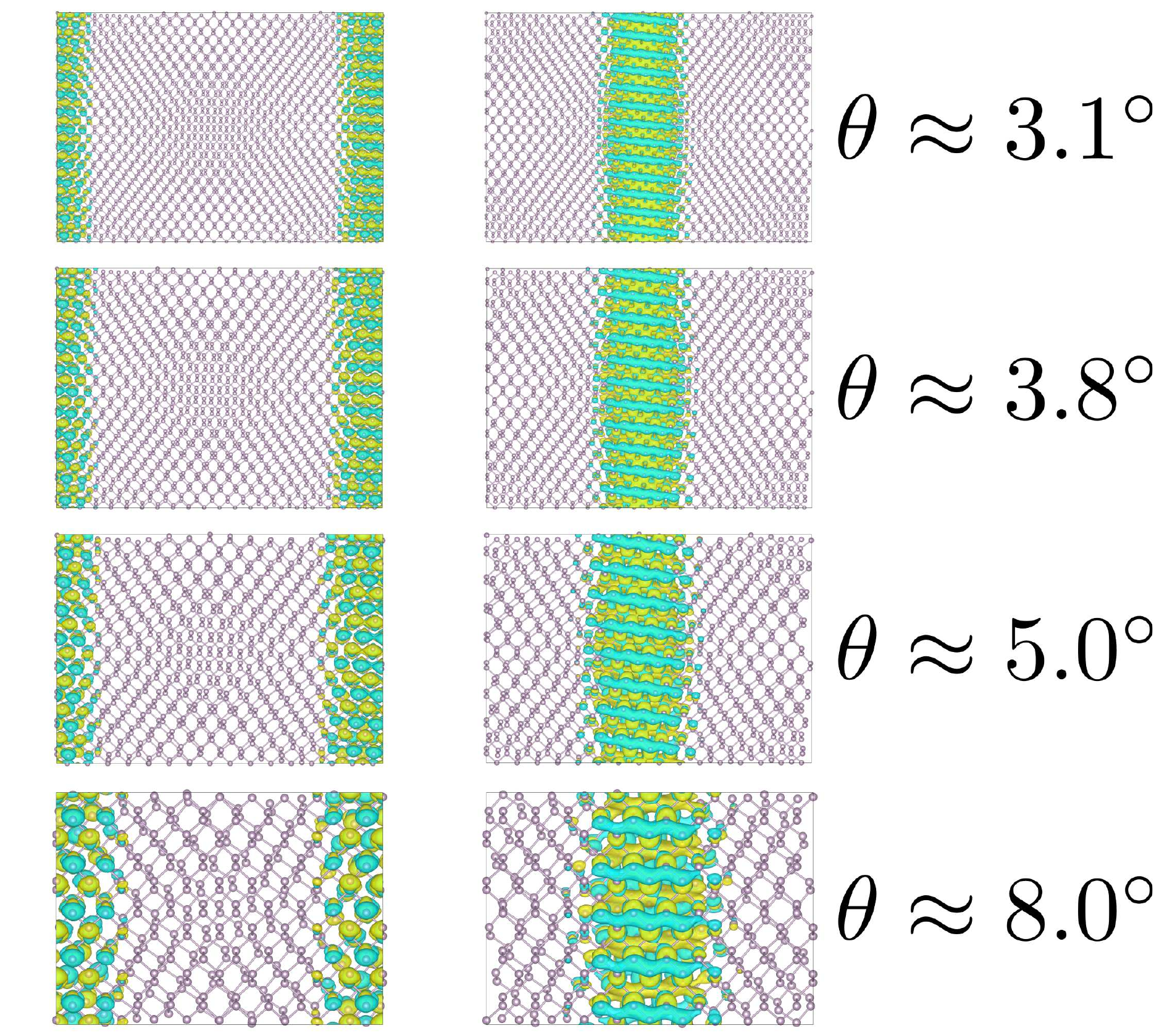}
	\caption{The VBM (left) and CBM (right) wave-function isosurfaces along the $z$-axis for each of the twisted phosphorene bilayer. The blue and yellow colors denote wave-function phase. All orbitals are highly localized in the $x$-direction. The VBM states bridge the AA--AB stacking regions, whereas the CBM states bridge the AA'-AB' stacking regions. As the twisting angle gets smaller, the unit cell increases and the impurity states become more separated.}
	\label{orb_twist}
\end{figure}

We now turn to the investigation of QP states in bilayers with decreasing twisting angle. The largest system investigated contains $2,708$ atoms (corresponding to $13,540$ valence electrons) and has a twist angle of $\theta\approx 3.1^\circ$. Note that conventional many-body calculations scale too steeply with the number of electrons and, hence, cannot be applied to such large systems. 	
As shown in Section~\ref{sec:relax} and in Figs.~\ref{fig:T8_relax} and \ref{orb_twist}, twisting leads to strong localization of the band edge states. Although a twisted structure has regions with characteristic stacking order (Figure~\ref{figure_structure}), it is not meaningful to explore stacking patterns independently. Rather, the twisted structure has to be considered as a whole. 

In order to investigate the behavior of the band edge states, we first inspect the energy-momentum characteristics of the eigenvectors $\phi$. The projector-based energy-momentum analysis (PEMA) is equivalent to the band unfolding for supercells constructed from ideally periodic unit cells \cite{popescu2012extracting,huang2014general}. We apply band unfolding to the twisted bilayer cell by projection onto the Brillouin zone of a single unit cell of a $0^\circ$-bilayer (detailed below). Note that this approach is fundamentally different from plotting the band structure within a Brillouin zone of the twisted cell \cite{kang2017moire}, which obfuscates the distinction between localized states and regular bands.

To perform the PEMA, eigenstates are transformed from a real-space grid to the $k$-space. Hence, $\phi_j$ is represented by a plane wave expansion:
\begin{equation}\label{FTphi}
\left| \phi_j \right\rangle = \sum_{\bg} C_j(\bg) e^{i\bg\cdot \bldr},
\end{equation}
where $\bg$ represents the reciprocal lattice vectors of the supercell. Note our use of real-space grid and supercells is equivalent to $\Gamma$ point sampling of the Brillouin zone. The band structure unfolding starts with computing an expectation value \cite{popescu2012extracting,huang2014general}:
\begin{equation}
P_j({\bf k}) \equiv \sum_n \left| \left\langle \phi_j \middle| {\bf k}, n \right\rangle \right|^2,
\end{equation}
where $\left | {\bf k}, n \middle \rangle \middle \langle {\bf k}, n \right |$ is a projector on a ${\bf k}$-dependent state in the first Brillouin zone of a \emph{single} unit cell. Here, ${\bf k}$ is a vector in the reciprocal space and $n$ is the band index. In the plane wave representation, the expectation value is computed directly from the Fourier components in Eq.~\ref{FTphi}:
\begin{equation}\label{proj_momentum}
P_j({\bf k}) = \sum_{\bf g} \left| C_j({\bf g} + {\bf k}) \right|^2,
\end{equation}
where ${\bf g}$ is the reciprocal lattice vectors of a single unit cell. Finally, the band structure is plotted as a function of ${\bf k}$ and QP energy of the $j^{\rm th}$ state. 
		
\begin{figure}
	\centering
	\includegraphics[width=0.75\textwidth]{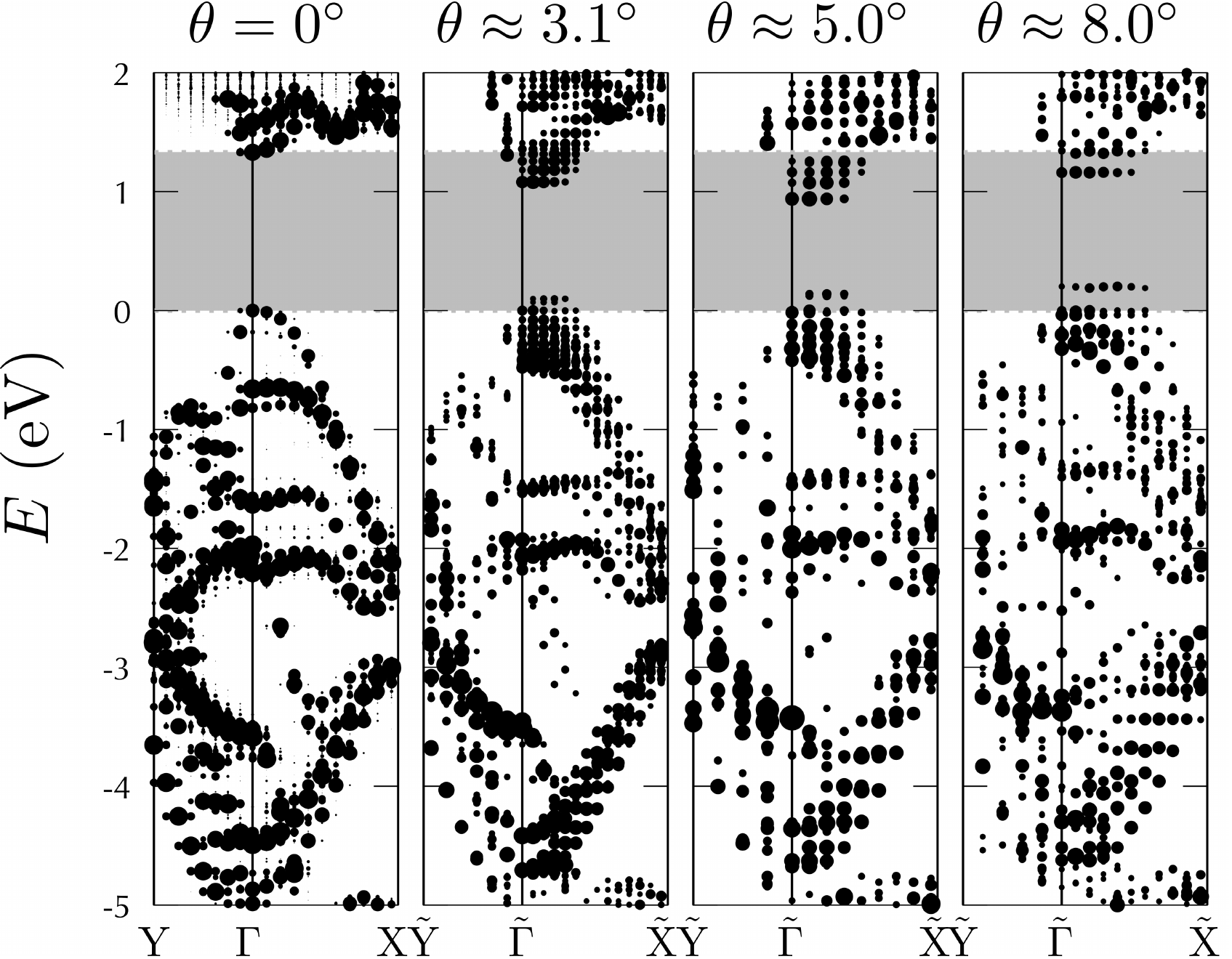}
	\caption{The QP band structures are computed by the PEMA discussed in Section~\ref{sec:res_twist}. Individual panels show the energy-momentum plots for various twisting angles $\theta$ along the ${\rm Y} \to \Gamma \to {\rm X}$ path of the Brillouin zone of a $0^\circ$-bilayer. The zero energy is defined as the maximum at ${\bf k} = 0$ that has the highest energy in the occupied subspace. The point size is proportional to $\log P({\bf k})$ (cf. Eq.~\ref{FTphi}). The paths in the Brillouin zones of the twisted cells sample corresponding $k$-vectors, but they are labeled $\tilde{{\rm Y}}$, $\tilde{\Gamma}$, and $\tilde{{\rm X}}$ as they do not correspond to the true critical points.}
	\label{BSTR}
\end{figure}
		
For perfectly periodic systems, $P_j({\bf k})$ is composed of $\delta$-functions and is fully equivalent to the band structure computed using conventional $k$ point sampling. This is illustrated for the ground state KS DFT results in Appendix B. Figure~\ref{BSTR} shows the $\bar\Delta GW_0$ band structure of the $0^\circ$-bilayer unfolded along the ${\rm Y}\to\Gamma$ and $\Gamma\to{\rm X}$ directions in the Brillouin zone of a single unit cell. The occupied and unoccupied states are shifted in energy to reproduce the QP band gap. The zero energy is defined as the maximum at ${\bf k} = 0$ that has the highest energy in the occupied subspace. For the $0^\circ$-bilayer, we see that the maxima in $P_j({\bf k})$ form valence and conduction bands. Slight numerical noise leads to partial smearing of the bands; nevertheless, the band structure is clearly visible.
		
With the exception of the $2\times2$ supercell for the $\theta\approx8.0^\circ$ system, we investigate merely single twisted unit cells (shown in Figure~\ref{orb_twist}). Hence, the electronic states cannot be unfolded in the strict sense. However, it is possible to perform a momentum analysis using Eqs.~\ref{FTphi} and \ref{proj_momentum} with ${\bf g}$-vectors of a bilayer unit cell with $\theta = 0^\circ$. Figure~\ref{BSTR} shows the energy-momentum plots for $\theta \approx 8.0^\circ$, $5.0^\circ$, and $3.1^\circ$. The energies of the bands are provided along the same $k$-vector path as for the bilayer. However, the horizontal axis is labeled as $\tilde{\rm Y}$, $\tilde\Gamma$, and $\tilde{\rm X}$ since the special $k$-points do not represent the critical points of the {\emph{true}} Brillouin zone. 
		
In Figure~\ref{BSTR}, we see that twisting causes energy splitting. Nevertheless, it is possible to observe individual valence bands for energies $<-1.5$~eV below the band edge. The deep valence states are localized mainly on individual monolayers, and they are thus less sensitive to twisting compared the states near the VBM. Still, large twisting angles lead to bandwidth reduction of all valence states. The arrangement of individual monolayers becomes more coherent for small $\theta$; the areas of individual stacking regions increase, and the average strain in the upper layer decreases (cf. Table~\ref{table_gaps}). As a result, the plot of $\theta\approx 3.1^\circ$ system is close to the band structure of a $0^\circ$-bilayer. 
		
More importantly, Figure~\ref{BSTR} captures the twist-induced localization. The Moir\'{e} impurity states appear in the gray-shaded area with minimal energy dispersion. Two valence impurities (cf.~Figure~\ref{fig:T8_relax}) have no intensity at ${\bf k}=0$. Hence, they appear just above the zero energy. Furthermore, they are separated from the valence region by an energy gap that decreases from $0.19\pm0.03$ to $0.06\pm0.03$~eV for angles between $8.0^\circ$ and $3.1^\circ$. In contrast, the conduction states are dragged to lower energies for small $\theta$; the empty Moir\'{e} impurity states are thus energetically close to the rest of the unoccupied states.
		
In Table~\ref{table_gaps}, we list the KS DFT eigenvalue gaps, and the QP gaps as predicted by $G_0W_0$ and by partially self-consistent $\bar\Delta GW_0$ approaches. While KS DFT predicts that $E_g$ decreases monotonically with $\theta$, the many-body methods show a different scenario. For the angles between $5.0^\circ$ and $3.1^\circ$ the band gap rises by $\sim0.16$~eV, which is eight times higher than the stochastic uncertainty and roughly twice as big as the estimated error due to the supercell size convergence discussed in Section~\ref{sec:convergence}.

\Table{\label{table_gaps} Table of results for all of the systems considered. $\bar{\epsilon}$ is the average of relative strains in the x and y direction between the top and bottom monolayers. Experimental and theoretical values of $E_g$ are given in the reference column with appropriate references in square brackets. All systems considered are planar besides monolayer and bilayer, which are fully relaxed.}
\br
&&&&\centre{4}{$E_g$ (eV)}\\
&&&&\crule{4}\\
\ns
&$\bar{\epsilon}$	& $a\times b$ (\AA)		&&PBE&$G_0W_0$				&$\bar{\Delta} GW_0$&Reference	\\
\mr
monolayer							&0\%				&$3.310\times4.636$		&&0.94&$2.07^{*}\pm0.03$			&$2.21^*\pm0.03$	&$2.0-2.08$ \cite{tran2014layer,liang2014electronic,qiu2017environmental}\\
bilayer							&0\%				&$3.300\times4.638$		&&0.62&$1.32^*\pm0.02$				&$1.37^*\pm0.03$	&$1.22-1.32$ \cite{tran2014layer,qiu2017environmental}\\
$\theta=8.0^\circ$	&0.98\%			&$33.00\times23.19$		&&0.37&$0.91^\star\pm0.03$		&$0.96^\star\pm0.03$&\\
$\theta=5.0^\circ$	&0.39\%			&$52.80\times37.10$		&&0.32&$0.79\pm0.02$				&$0.82\pm0.02$&\\
$\theta=3.8^\circ$	&0.22\%			&$69.30\times51.02$		&&0.28&$0.83\pm0.02$				&$0.86\pm0.02$&\\
$\theta=3.1^\circ$	&0.15\%			&$85.80\times60.29$		&&0.26&$0.90\pm0.03$				&$0.98\pm0.03$&\\
\br
\end{tabular}
\item[] $^{\rm *}$ estimated from interpolation shown in Figure~\ref{fig:converge}
\item[] $^{\star}$ calculations based on $2\times2$ supercell with dimensions $66.00\times46.38$ \AA
\end{indented}
\end{table}

The differences between the mean-field DFT and the $GW$ results stem from the treatment of electron-electron interactions. In DFT, the xc term is represented by a local potential, which changes only negligibly (it decreases linearly with $\theta$ with a slope of $\sim0.01$~eV/$^\circ$). The KS eigenvalue response is governed by the changes from the external and Hartree terms. Together, the two electrostatic potentials tend to destabilize the occupied states and lead to band gap closing. 

In the many-body calculations, the changes in the Hartree and external potentials are compensated by the non-local self-energy. For the valence impurity state, $\Sigma$ is dominated by exchange interaction that grows more negative with orbital localization. For the valence impurities, the stabilizing exchange term is roughly 80-100 times larger than the polarization contribution, which shifts QP energies up. Since the valence Moir\'{e} states become more localized with decreasing $\theta$, their energy decreases as they are governed by $\Sigma_x$.

The conduction states behave differently because the exchange interaction is much weaker for unoccupied states. Furthermore, the $\Sigma_x$ expectation value is diminished in bilayer phosphorene due to small orbital overlaps in Eq.~\ref{eq:X}. This is related to the distinct spatial distribution of the occupied and unoccupied states: the conduction impurity orbital is predominantly located in the interlayer region, while the occupied states spread over the monolayers (Figure~\ref{fig:T8_relax}). In contrast to the valence states, the polarization self-energy is as large as the exchange term (due to enhanced polarizability term in Eq.~\ref{eq:P}), and it shifts the QP energies down. States with a high orbital density in the interlayer region are strongly coupled to polarization modes, which are also responsible for the van der Waals bonding. Dynamical and non-local $\Sigma_P$ captures such effects in the QP energies (unlike the underlying KS DFT). In general, we observe stabilization of the conduction states through the polarization term, i.e., the empty states are shifted to lower energies as seen in Figure~\ref{BSTR} for $\theta\approx 5.0^\circ$.

The band gap variation in the twisted bilayer is thus governed by the interplay of non-local electron-electron interactions, which distinctly affect occupied and unoccupied Moir\'{e} impurities. The valence states are stabilized by the exchange interaction. In contrast, the conduction states are subject to a competition of exchange and polarization effects which have similar strength but shift the QP energies in opposite directions. The interplay between exchange and polarization interactions is common for all van der Waals (hetero)structures. Hence, we expect that the same scenario applies to a wide class of systems, in which the non-local many-body effects may be tuned by variable twisting angle.

\section{Conclusions}

In this work, we investigate the QP energies of the Moir\'{e} impurity states in twisted bilayer phosphorene using many-body perturbation theory. To perform such calculations, we introduce a new implementation of the stochastic $GW$ approach suitable for 2D-periodic systems. Similar to 3D solids, the stochastic fluctuations converge faster for large systems. The comparison between $0^\circ$ and twisted bilayers reveals that the stochastic error is increased by $\sim 30\%$ due to a twist-induced localization. Hence, the convergence of the stochastic error is slower and around twice as many random vectors are needed. The current stochastic formulation for 2D systems allows treatment of supercells with thousands of atoms. We demonstrate these capabilities on twisted bilayer phosphorene systems with up to $2,708$ phosphorus atoms, i.e., with $13,540$ valence electrons. The new implementation is general, and it provides a powerful tool to study Moir\'{e} impurities.

The structures of twisted bilayer phosphorene are characterized by areas with distinct stacking order, which causes variation in the local geometry. The fully relaxed structures are corrugated, and the interlayer distance is not uniform. The conduction impurity states are more strongly affected by relaxation because they are localized mostly in the interlayer region. However, corrugation leads to relatively small changes in the QP energies compared to the effect of bare twisting. We thus focus on QPs in planar bilayers induced purely by changes in the twisting angle $\theta$.

To investigate the twist-induced in-gap states, we employ a projector-based energy-momentum analysis method which clearly illustrates the formation of strongly localized orbitals. Both valence and conduction impurities appear above/below the band edge energies of a $0^\circ$-bilayer. The occupied Moir\'{e} impurities are well-separated from the rest of the valence bands, but this separation decreases with small $\theta$. The unoccupied impurities are close to the conduction bands, which are pushed to lower energies with the decreasing twisting angle. 

The behavior of the Moir\'{e} states and the size of the band gap are governed by an interplay between local (ionic and Hartree) and non-local (electron-electron) interactions. Twist-induced localization affects the valence and conduction states distinctly. The former is stabilized by electron localization, while the dynamical correlation strongly influences the latter.

Twisting thus introduces a unique way to modify the electron-electron interactions with high precision. Twisted phosphorene bilayers represent one class of 2D system with Moir\'{e} states, but the described mechanism of impurity energy (de)stabilization is general and applicable to other low-dimensional  van der Waals (hetero)structures.

\ack
This work was supported by the NSF through the Materials Research Science and Engineering Centers (MRSEC) Program of the NSF through Grant No. DMR-1720256 (Seed Program). The calculations were performed as part of the XSEDE~\cite{XSEDE} computational Project No. TG-CHE180051. Use was made of computational facilities purchased with funds from the National Science Foundation (CNS-1725797) and administered by the Center for Scientific Computing (CSC). The CSC is supported by the California NanoSystems Institute and the Materials Research Science and Engineering Center (MRSEC; NSF DMR 1720256) at UC Santa Barbara. Jacob Brooks was supported by the UCSB Edison Summer Research Program which is funded by Edison International.\\

\section*{Appendix A: }
\renewcommand{\thetable}{A\arabic{table}}
\setcounter{table}{0}

\begin{table}[H]
\caption{\label{table_re_distance} The interlayer distances associated with each stacking region. Each distance is taken at a local extremum  (AA' and AB regions) or a saddle point (AA and AB' regions).}
\begin{indented}
	\item[]\begin{tabular}{@{}ll}
		\br
		Stacking Region&$\bar{d}$ (\AA)\\
		\mr
		AA&$3.47$\\
		AA'&$3.72$\\
		AB&$3.28$\\
		AB'&$3.51$\\
		\br
	\end{tabular}
\end{indented}
\end{table}
\section*{Appendix B:}
\renewcommand{\thefigure}{B\arabic{figure}}
\setcounter{figure}{0}

\begin{figure}[H]
\centering
\includegraphics[width=0.5\textwidth]{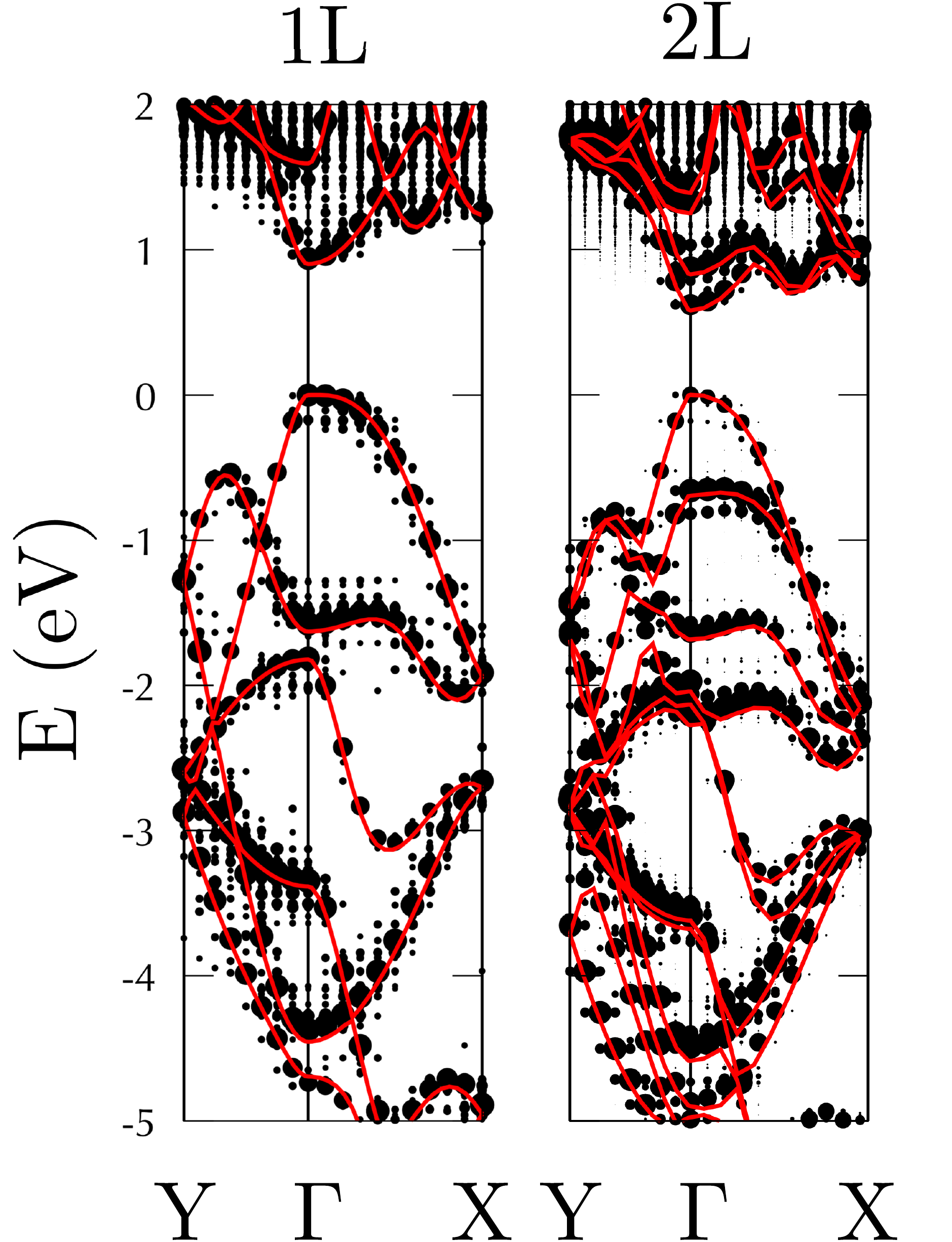}
\caption{Band unfolding (black dots) compared with the band structure computed with ${\bf k}$-point sampling computed in Quantum Espresso code (red lines) for the monolayer (1L) and bilayer (2L) systems. The zero energy is defined as the maximum at ${\bf k} = 0$ that has the highest energy in the occupied subspace. The point size is proportional to $\log P({\bf k})$ (cf. Eq.~\ref{FTphi}). Energy states are shown for the ${\bf k}$-vectors in the first Brillouin zone.}\label{fig:band_qe_comp}
\end{figure}

\bibliography{TST}
\end{document}